\documentclass{article}
\usepackage{spconf,amsmath,graphicx,hyperref}
\usepackage{multirow} 
\usepackage{multicol}   
\usepackage{booktabs}  
\usepackage{amssymb}
\usepackage[all]{hypcap}

\title{Bone-conduction Guided Multimodal Speech Enhancement with Conditional Diffusion Models}
\name{Sina Khanagha, Bunlong Lay, Timo Gerkmann}
\address{Signal Processing Group, University of Hamburg, Germany}

\begin{document}
\maketitle
\begin{abstract}
Single-channel speech enhancement models face significant performance degradation in extremely noisy environments. While prior work has shown that complementary bone-conducted speech can guide enhancement, effective integration of this noise-immune modality remains a challenge. This paper introduces a novel multimodal speech enhancement framework that integrates bone-conduction sensors with air-conducted microphones using a conditional diffusion model. Our proposed model significantly outperforms previously established multimodal techniques and a powerful diffusion-based single-modal baseline across a wide range of acoustic conditions.

\end{abstract}
\begin{keywords}
Bone-conducted speech, multimodal speech enhancement, conditional diffusion models, speech enhancement, score-based generative models
\end{keywords}
\section{Introduction}
\label{sec:intro}

Conventional speech enhancement models aim to reconstruct the original clean speech signal from the corrupted noisy mixture. Recent deep neural network (DNN) based speech enhancement models can be categorized into predictive and generative models \cite{10095258}. Predictive models aim to directly estimate the clean speech from the noisy mixture whereas the generative models employ strategies to utilize the noisy mixture as a conditioning signal to guide a stochastic generation process in order to create plausible clean speech samples that resemble the original clean speech.

Naturally, the denoising performance of both types depends on the intensity of the noise. At very low signal-to-noise ratios (SNRs) the models often fail to output adequate results. This motivates the idea of moving beyond the single-modality speech enhancement models and utilizing auxiliary modalities that are immune to acoustic noise such as visual cues \cite{10363042}, or bone-conducted speech \cite{10778414, 10888094, WANG2022109058, 9903559, 9112325, 10889416}.

Bone-conducted speech is commonly recorded with an accelerometer by tracking the local vibrations of human skull which in turn originates from the vibrations of the vocal tract. Although by nature this modality of speech is immune against acoustic noise, it suffers from limited bandwidth and poor intelligibility. Therefore, a plausible idea is to simultaneously employ both modalities to construct multimodal speech enhancement models that are able to take advantage of both the acoustic noise immunity of the band-limited bone-conducted speech and the high-frequency content of the air-conducted speech.

Most of the recent bone-conduction guided multimodal speech enhancement models utilize a U-Net like backbone \cite{10778414, 9903559, 9112325} and make an effort to introduce the auxiliary bone-conducted signal into the denoising process with various strategies. The most prominent method is to either directly fuse the conditioning signal with the noisy mixture at the input \cite{9903559, 9112325} or to implement an attention based feature pre-processing layer to construct a fused input for the network \cite{10888094, 9903559}. Furthermore, some of the methods utilize separate downsampling encoders for each modality and perform the feature fusion steps on the embeddings of each encoder before the bottleneck layer \cite{10778414, 9112325}. Li et al. \cite{10778414} also introduced a contrastive learning loss as a regularizer term for the embeddings of each encoder in order to reduce the embedding distance of the two modalities which marginally improves the enhancement performance of the model. Duan et al. \cite{10800089} also utilizes a diffusion process as a complimentary generation step for bandwidth extension of the bone-conducted speech. It is important to note however that this pursues a different goal than multimodal speech enhancement.

In this paper, we introduce the Bone-conduction Conditional Diffusion Model (BCDM), a complex-domain diffusion model for multimodal speech enhancement. To effectively integrate bone-conduction sensor data, we propose and compare two distinct conditioning strategies for a U-Net denoiser: a direct input concatenation (IC) method and a decoder conditioning (DC) feature injection approach. Extensive experiments demonstrate that BCDM substantially outperforms both single-modality diffusion and state-of-the-art multimodal baselines across low and high SNR conditions. To the best of our knowledge, this is the first work to apply a conditional diffusion framework to bone-conduction guided speech enhancement.

\section{Method}
\label{sec:format}

The fundamental idea behind recent generative diffusion models is to gradually add noise to the samples until the sample transforms into pure noise in the forward process and then training a DNN to reverse this process and generate high fidelity samples with respect to our original data distribution. The general forward process is expressed as the following stochastic differential equation (SDE) \cite{song2021scorebased}:

\begin{equation}
    dx_t = f(x_t, t)\,dt + g(t)\,dW_t, \quad t \in [0, T],
\label{eq:general-sde}
\end{equation}
where~$x_t$ denotes the process state at time step $t$ and $W_t$ denotes a standard Wiener process. Moreover, $f(x_t, t)$ and $g(t)$ represent the \textit{drift} and \textit{diffusion} coefficients, respectively. The drift coefficient dictates the deterministic evolution and overall direction of the process whereas the diffusion coefficient controls the stochastic part of the process by determining the magnitude of random fluctuations added at each step.

Speech enhancement can be expressed as a conditional generation task with the goal of generating clean speech distribution samples conditioned on the noisy mixture \cite{9746901, 10149431}. In our work, we utilize the same conditioning method proposed by Richter et al. \cite{10149431} in which the drift coefficient is modified as:
\begin{equation}
    f(x_t, y) := \gamma (y-x_t) ,
\end{equation}

where $y$ denotes the noisy air-conducted mixture and $\gamma$ denotes the \textit{stiffness} constant controlling the mean evolution of the process from clean speech $x_0$ to $y$. Essentially, instead of a transition of samples to zero-mean noise, we move the mean towards the noisy mixture distribution. Furthermore, similar to the variance exploding method of Song et al. \cite{song2021scorebased}, we define the diffusion coefficient as:

\begin{equation}
g(t) := \sigma_{\min} \left( \frac{\sigma_{\max}}{\sigma_{\min}} \right)^t 
\sqrt{ 2 \log \left( \frac{\sigma_{\max}}{\sigma_{\min}} \right) },
\end{equation}

Where $\sigma_{\max}$ and $\sigma_{\min}$ parameters bound the noise schedule of the Wiener process. The choice of these parameters play a crucial role in denoising capabilities of the model as it has been shown that having a higher $\sigma_{\max}$ results in a superior environmental noise removal at the cost of losing more speech components \cite{lay24_interspeech}. In our work we opted for $\sigma_{\min} = 0.05$ and $\sigma_{\max} = 0.5$ as it provides a reasonable trade-off between noise removal and speech preservation.

The associated reverse SDE for \eqref{eq:general-sde} can be derived as:
\begin{equation}
    dx_t = [-f(x_t, y) + g(t)^2 \nabla_{x_t} \log{p_t}(x_t|y)]dt+g(t)d\overline{w}
\label{eq:reverse-sde}
\end{equation}

The \textit{score} term $\nabla_{x_t} \log{p_t}(x_t|y)$ is intractable but can be estimated with a DNN, resulting in a score \emph{model}. We denote our score model with $s_\theta(x_t, y, y_c, t)$ which receives the current process state $x_t$, noisy air-conducted speech $y$, the conditional bone-conducted speech signal $y_c$, and diffusion time step $t$ as inputs, with $\theta$ representing the parameters of the DNN. By swapping  the score model in \eqref{eq:reverse-sde} we have:
\begin{equation}
    dx_t = [-f(x_t, y) + g(t)^2 s_\theta(x_t, y, y_c, t)]dt+g(t)d\overline{w}
\end{equation}

Which during inference can be solved by an ODE solver for deterministic results or a stochastic SDE solver for sampling-based solutions. In this work we have utilized the predictor-corrector (PC) sampler strategy proposed by Song et al. \cite{song2021scorebased} as our SDE solver. 

Moreover, the objective function for score model training is the score matching objective described as:
\begin{equation}
\mathbb{E}\left[ \| s_\theta(x_t, y, y_c, t) - \nabla_{x_t} \log p_{0t}(x_t \mid x_0, y) \|_2^2 \right],
\end{equation}
where the expectation $\mathbb{E}$ is taken over $(t, x_t)$ conditioned on $(x_0,y, y_c)$ with $t \sim \mathcal{U}(0,1)$. In addition, $\nabla_{x_t} \log p_{0t}(x_t \mid x_0, y)$ can be efficiently calculated at each randomly sampled step t according to \cite{10149431}.

\begin{figure*}[htbp]
    \centering
    \includegraphics[width=0.99\textwidth]{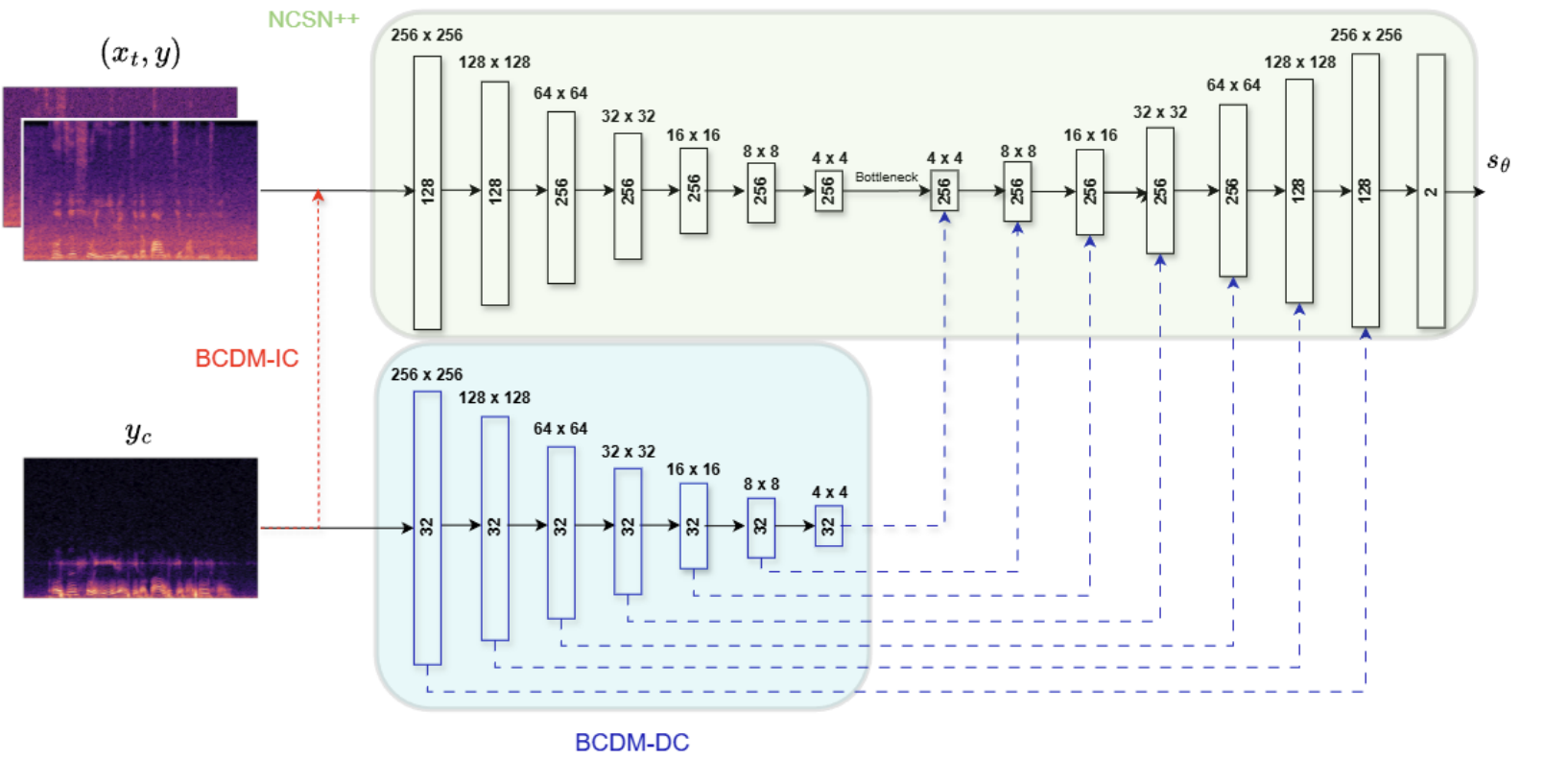}
    \caption{An overview of two IC (red) and DC (blue) conditioning strategies for the large variant of BCDM. The skip connections and progressive layers of the NCSN++ have been excluded from this diagram for clarity improvement}
    \label{diagram}
\end{figure*}

\section{Conditioning Strategies}
In this section we introduce our two conditioning strategies for inserting the bone-conducted speech signal into the score model.
The fundamental backbone architecture of the score model in our experiments is a modified version of NCSN++ \cite{song2021scorebased} proposed by \cite{10149431} which is a multi-resolution U-Net type DNN. Two configurations of the backbone have been considered where the smaller one has half the number of feature maps at each down or upsampling layers and the depth of the ResNet blocks are reduced to 1. The large configuration is referred to with the suffix "-L" and the smaller one with "-S", respectively. An overview of the conditioning methods and layer configurations of the large variant is presented in \autoref{diagram}.

\subsection{Input Concatenation (IC)}
For the first method, we concatenate the time-aligned bone-conducted speech spectrogram with both the noisy air-conducted speech spectrogram and the current process state spectrogram along the channel dimension. The advantage of this method lies in its simplicity and minimal number of added parameters to the network. Additionally, due to the similarity of the two modalities, theoretically having a mutual encoder with the same architecture should not drastically limit the learning capabilities. We denote this method with BCDM-IC.

\subsection{Decoder Conditioning (DC)}
It can also be beneficial to utilize separate encoders for each modality to increase the network's capability to extract nuance information from each modality individually instead of dealing with a fused input feature immediately at the initial layers. Also, it has been demonstrated that injecting the conditioning feature maps into the upsampling layers of the score model can result in significant performance gains \cite{kim24o_interspeech}. Such conditioning resembles the typical skip connections in the U-Net backbone.
Hence, we designed a separate encoder for the bone-conducted speech that injects the resulting feature maps into decoder's upsampling layers at each corresponding resolution. The conditioning encoder downsampling blocks are constructed from ResNet type blocks proposed in the BigGAN \cite{brock2018large} model which are also conditioned on the time embeddings of the diffusion process. Finally, after concatenating the feature maps from skip connections, condition encoder, and the previous layer's output feature maps, we have utilized 1x1 Conv2D layers to reduce the number of feature maps to match the original NCSN++ decoder feature maps in order to avoid a significant increase in parameters. We denote this method with BCDM-DC. Note that in both large and small variants of this method, configuration of the conditioning encoder is identical.

\begin{table*}[ht]
\centering
\caption{Comparison of models across SNR levels with metrics POLQA, PESQ, and ESTOI on 3741 test utterances. The average POLQA, PESQ and ESTOI values for bone-conducted speech are 1.89 ± 0.45, 1.43 ± 0.25, and 0.58 ± 0.09, respectively. For all diffusion-based models we used N=60 denoising reverse steps. \textsuperscript{*}SGMSE+ is the only single-modality model.}
\label{tab:table-1}
\small
\resizebox{\textwidth}{!}{
\begin{tabular}{c|c|ccccccccc}
\toprule
\textbf{SNR (dB)} & \textbf{Metric} & 
\begin{tabular}[c]{@{}c@{}}Noisy \\ Mixture\end{tabular} & 
\begin{tabular}[c]{@{}c@{}}SGMSE+* \\ (65.6M)\end{tabular} & 
\begin{tabular}[c]{@{}c@{}}FCN-LF \\ (0.26M)\end{tabular} & 
\begin{tabular}[c]{@{}c@{}}DCCRN \\ (13.8M)\end{tabular} & 
\begin{tabular}[c]{@{}c@{}}BiNet \\ (27.3M)\end{tabular} & 
\begin{tabular}[c]{@{}c@{}}BCDM-IC-S \\ (11.7M)\end{tabular} & 
\begin{tabular}[c]{@{}c@{}}BCDM-DC-S \\ (12.3M)\end{tabular} & 
\begin{tabular}[c]{@{}c@{}}BCDM-IC-L \\ (65.6M)\end{tabular} & 
\begin{tabular}[c]{@{}c@{}}BCDM-DC-L \\ (67.4M)\end{tabular} \\
\midrule
\multirow{3}{*}{-10} 
 & POLQA & 1.09 ± 0.07 & 1.30 ± 0.35 & 1.33 ± 0.25 & 1.93 ± 0.38 & 2.35 ± 0.40 & 2.36 ± 0.45 & 2.31 ± 0.48 & 2.37 ± 0.45 & \textbf{2.44 ± 0.46} \\
 & PESQ   & 1.08 ± 0.21 & 1.15 ± 0.17 & 1.08 ± 0.06 & 1.40 ± 0.20 & 1.80 ± 0.26 & 1.86 ± 0.39 & 1.92 ± 0.38 & 1.95 ± 0.36 & \textbf{2.02 ± 0.40} \\
 & ESTOI  & 0.21 ± 0.11 & 0.36 ± 0.19 & 0.38 ± 0.12 & 0.63 ± 0.09 &  0.70 ± 0.08 & 0.75 ± 0.08 & 0.74 ± 0.08 & \textbf{0.76 ± 0.08} & \textbf{0.76 ± 0.08}\\

\midrule
\multirow{3}{*}{-5} 
 & POLQA  & 1.14 ± 0.15 & 1.81 ± 0.51 & 1.48 ± 0.30 & 2.18 ± 0.40 & 2.46 ± 0.40 & 2.51 ± 0.49 & 2.51 ± 0.52 & 2.61 ± 0.48 & \textbf{2.67 ± 0.50} \\
 & PESQ   & 1.06 ± 0.07 & 1.42 ± 0.33 & 1.15 ± 0.13 & 1.60 ± 0.27 & 1.93 ± 0.30 & 2.02 ± 0.44 & 2.09 ± 0.44 & 2.19 ± 0.42 & \textbf{2.24 ± 0.45} \\
 & ESTOI  & 0.34 ± 0.12 & 0.62 ± 0.15 & 0.49 ± 0.13 & 0.70 ± 0.08 & 0.74 ± 0.08 & 0.78 ± 0.08 & 0.77 ± 0.08 & \textbf{0.80 ± 0.08} & 0.79 ± 0.08 \\
\midrule
\multirow{3}{*}{0} 
 & POLQA  & 1.28 ± 0.28  & 2.37 ± 0.54 & 1.68 ± 0.37 & 2.45 ± 0.39 & 2.54 ± 0.41 & 2.73 ± 0.53 & 2.73 ± 0.54 & 2.86 ± 0.49 & \textbf{2.93 ± 0.52}\\
 & PESQ   & 1.08 ± 0.06 & 1.84 ± 0.44 & 1.29 ± 0.21 & 1.83 ± 0.34 & 2.05 ± 0.33 & 2.22 ± 0.48 & 2.30 ± 0.48 & 2.42 ± 0.45 &  \textbf{2.49 ± 0.48} \\
 & ESTOI  & 0.50 ± 0.13 & 0.76 ± 0.10 & 0.62 ± 0.12 & 0.77 ± 0.08 & 0.77 ± 0.07 & 0.82 ± 0.08 & 0.81 ± 0.08 & \textbf{0.84 ± 0.07} & 0.83 ± 0.08\\
\midrule
\multirow{3}{*}{5} 
 & POLQA  & 1.55 ± 0.40 & 2.83 ± 0.50 & 1.94 ± 0.41 & 2.73 ± 0.40 & 2.62 ± 0.43 & 3.00 ± 0.54 & 3.00 ± 0.55 & 3.12 ± 0.49 & \textbf{3.20 ± 0.50}\\
 & PESQ   & 1.17 ± 0.12& 2.30 ± 0.47 & 1.50 ± 0.27 & 2.08 ± 0.39 & 2.14 ± 0.35 & 2.49 ± 0.51 & 2.54 ± 0.51 & 2.67 ± 0.46 & \textbf{2.74 ± 0.49} \\
 & ESTOI  &0.65 ± 0.12 & 0.85 ± 0.07 & 0.73 ± 0.10 & 0.83 ± 0.07 & 0.80 ± 0.06 & 0.87 ± 0.07 & 0.86 ± 0.07 & \textbf{0.88 ± 0.06} & 0.87 ± 0.06\\
\midrule
\multirow{3}{*}{15} 
 & POLQA & 2.42 ± 0.53 & 3.55 ± 0.44 & 2.49 ± 0.37 & 3.25 ± 0.46 & 2.78 ± 0.46 & 3.59 ± 0.51 & 3.53 ± 0.49 & 3.60 ± 0.46 & \textbf{3.70 ± 0.45} \\
 & PESQ   & 1.74 ± 0.33 & 3.08 ± 0.41 & 1.95 ± 0.31 & 2.61 ± 0.43 & 2.30 ± 0.38 & 3.10 ± 0.49 & 3.08 ± 0.47 & 3.16 ± 0.42 & \textbf{3.25 ± 0.43} \\
 & ESTOI  & 0.87 ± 0.07 & \textbf{0.94 ± 0.04} & 0.83 ± 0.07 & 0.92 ± 0.04 & 0.84 ± 0.05 & \textbf{0.94 ± 0.04} & 0.93 ± 0.04 & \textbf{0.94 ± 0.04} & \textbf{0.94 ± 0.04} \\
\bottomrule
\end{tabular}
}
\end{table*}

\section{Experimental Setup}

For our experiments, we utilized the ABCS dataset \footnote{\url{https://github.com/wangmou21/abcs}}
 \cite{WANG2022109058}, which contains 42 hours of time-aligned air-conducted and bone-conducted speech recordings. The dataset comprises utterances of 1–5 seconds in duration, collected from 100 Chinese speakers. Following the original setup, the dataset is divided into training, validation, and test subsets with 84, 8, and 8 speakers, respectively, and we adopted the same speaker partitions in our experiments. To create the noisy air-conducted mixtures for our training and testing, we used the CHiME3 dataset \cite{Barker2015_ASRU}. The SNR value for each training utterance has been uniformly sampled from -5 to 20 dB.

We extract complex-valued STFT representation using a window size of 510 and the hop length of 128, resulting in 256 frequency bins. The length is also trimmed to 256 frames to construct the 256 x 256 complex input of the score model. The batch size is set to 8 and we used Adam optimizer with the learning rate of $10^\text{-4}$. The exponential moving average of the DNN weights are also recorded with a decay rate of 0.999 for sampling \cite{10149431}.

We compare our results with three of the recently proposed bone-conduction guided multimodal speech enhancement models as our baselines: FCN-LF \cite{9112325}, DCCRN \cite{9903559}, and BiNet \cite{10778414}. We also included the single-modality SGMSE+ model \cite{10149431}, which shares the NCSN++ backbone with ours but lacks the conditioning additions.

The metrics used for comparison are extended short-time objective intelligibility (ESTOI) \cite{7539284}, wide-band perceptual evaluation of speech quality (PESQ) \cite{941023}, and the perceptual objective listening quality analysis (POLQA) \cite{polqa}. All of the models have been trained from scratch by us and the checkpoint with the highest average PESQ value for 20 randomly chosen utterances of validation speakers is chosen for comparisons.

\section{Results and Discussions}
\label{sec:majhead}
We created subsets with various noise intensities from the test speakers, totaling 3,741 samples per condition where the SNR value is randomly drawn according to Gaussian distributions centered at -10, -5, 0, 5, 15 with $\sigma = 1$. The results are presented in \autoref{tab:table-1}.

\subsection{Comparison Against Baseline Models}
\label{ssec:subhead}

As demonstrated in \autoref{tab:table-1}, all of our proposed BCDM models outperform the recently proposed models across all acoustic conditions. Although the BiNet model is competitive with our small models at low SNR conditions, it is evident that there is a clear under-utilization of the air-conducted modality in the BiNet model that is specially apparent at higher SNR where at 15 dB, our small variants outperform BiNet by as much as 0.81 POLQA, 0.8 PESQ and 0.1 ESTOI. The BiNet model essentially acts similarly to a bandwidth extension model rather than a multimodal speech enhancement which explains the relative low performance gain in higher SNRs. Conversely, the DCCRN model demonstrates better adaptiveness across different conditions but fails to properly utilize the bone-conducted modality which results in subpar performance at lower SNRs. Moreover, all of the multimodal baselines only manage to outperform the SGMSE+ single-modality model at extreme conditions where we can observe that even at 5 SNR, SGMSE+ outperforms other baselines. However, even at 15 SNR, BCDM large variants with comparable size comfortably outperform the SGMSE+ in terms of POLQA and PESQ while the small variants with approximately six times fewer parameters demonstrate almost identical performance across all metrics.

\subsection{Number of Reverse Steps}
\label{ssec:subsubhead}

\begin{figure}[t]
    \centering
    \includegraphics[width=0.75\columnwidth]{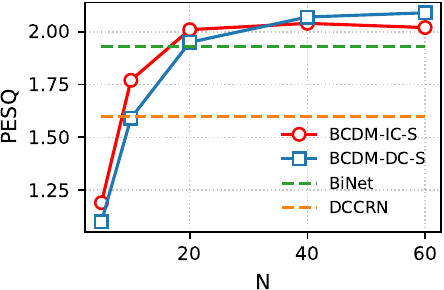}
    \caption{PESQ evolution of BCDM's small variants as a function of $N$ at -5 dB SNR}
    \label{fig:pesq_vs_n}
\end{figure}

Although our models exhibit superior performance, it is important to acknowledge that other baselines are predictive models which produce the output in a single function call whereas in diffusion frameworks several reverse steps N are required to produce adequate results. As demonstrated in \autoref{fig:pesq_vs_n}, BCDM small variants surpass DCCRN at around 10 steps and BiNet with 20 steps where each step corresponds to 2 score function calls in the PC sampling strategy. It is also noticeable that the IC strategy requires fewer steps to reach its peak performance compared to DC strategy.

\section{Conclusion}
In this work we introduced BCDM, a multimodal bone-conduction guided speech enhancement model. Our experiments demonstrate that BCDM significantly outperforms recent multimodal baselines and a strong single-modality baseline across all acoustic conditions. Furthermore, we proposed two conditioning strategies where our experiments suggest that the DC method offers a noticeable improvement in instrumental metrics at the cost of more reverse steps and more network parameters compared to the IC method.
\label{sec:print}

\bibliographystyle{IEEEtran}
\bibliography{refs}

\end{document}